\documentclass[12pt]{iopart}

\usepackage{graphicx}% Include figure files
\usepackage{dcolumn}% Align table columns on decimal point
\usepackage{bm}% bold math

\begin{document}

\title[Xiao \& Mortensen, Proposal of highly sensitive optofluidic ...]{Proposal of highly sensitive optofluidic sensors based on dispersive photonic crystal waveguides}

\author{Sanshui Xiao and Niels Asger Mortensen}

\address{MIC -- Department of Micro and Nanotechnology, NanoDTU,\\Technical
University of Denmark, DK-2800 Kongens Lyngby, Denmark.}
\ead{sanshui.xiao@mic.dtu.dk}
\begin{abstract}
Optofluidic sensors based on highly dispersive two-dimensional
photonic crystal waveguides are theoretically studied. Results
show that these structures are strongly sensitive to the
refractive index of the infiltrated liquid($n_l$), which is used
to tune dispersion of the photonic crystal waveguide. Waveguide
mode-gap edge shifts about 1.2~nm for $\delta n_l$=0.002. The
shifts can be explained well by band structure theory combined
with first-order perturbation theory. These devices are
potentially interesting for chemical sensing applications.

\end{abstract}

\pacs{07.07.Df, 42.79.Gn, 42.70.Qs} \vspace{2pc} \noindent{\it
Keywords}: Sensors, Optical waveguide, Photonic Crystal
\\ \submitto{\JOA, Nanometa 2007 special issue.}
\maketitle

\section{Introduction}
Optofluidics, the marriage of nano-photonics and micro-fluidics,
refers to a class of optical systems that integrate optical and
fluidic devices~\cite{Psaltis:2006}. Due to unique properties of
fluids, such integration provides a new way for dynamic
manipulation of optical properties and shows many potential
applications~\cite{Domachuk:2004,Erickson:2006,Galas:2005,Grillet:2004,Kurt:2005,GersborgHansen:2005,Li:2006,GersborgHansen:2006}.
One of the most exciting optofluidic applications is to realize
sensors, which can be used to detect, manipulate and sort cells,
virus and biomolecules in fluidics. Many optical sensors operate
by measuring the change in refractive index at the surface of the
sensor with the method of e.g. surface-plasmon resonance,
colorimetric resonances, interferometry in porous silicon.
However, these methods require large-area beams and relatively
large sensing area. Recently, Chow {\it et al.} demonstrated an
ultra compact sensor employing a two-dimensional photonic crystal
microcavity~\cite{Chow:2004}.

Photonic crystals (PhCs) are attractive optical materials for
controlling and manipulating the flow of
light~\cite{John:1987,Yab:1987,Joannopoulos:1995}. Because of the
unique light-confinement mechanism provided by the photonic
bandgap, they have attracted much attention recently. It has been
shown that they have many potential applications in
optoelectronics, such as high-quality-factor
filters~\cite{Xiao:2005}, low-threshold lasers~\cite{Park:2004},
optical switches, etc. In particular, PhCs are interesting for
optofluidics since they naturally have voids where fluids can be
injected~\cite{Xiao:2006b,Xiao:2006c}. If we introduce a defect in
a perfect PhC, optical properties of the defect can be easily
reconfigured by selectively filling specific voids with liquid.
PhC-based waveguides are highly dispersive, which has been used to
slow down the propagation velocity of light~\cite{Gersen:2005}. In
this paper, we will propose simple sensor structures based on
highly dispersive photonic crystal waveguides. These structures
are strongly sensitive to the refractive index of the liquid which
is used to tune dispersion of the PhC waveguide. In the following
we consider two different realizations of such a sensor.

\section{Sensor structures and discussion}

\begin{figure}[ht]
\centering\includegraphics[width=\columnwidth]{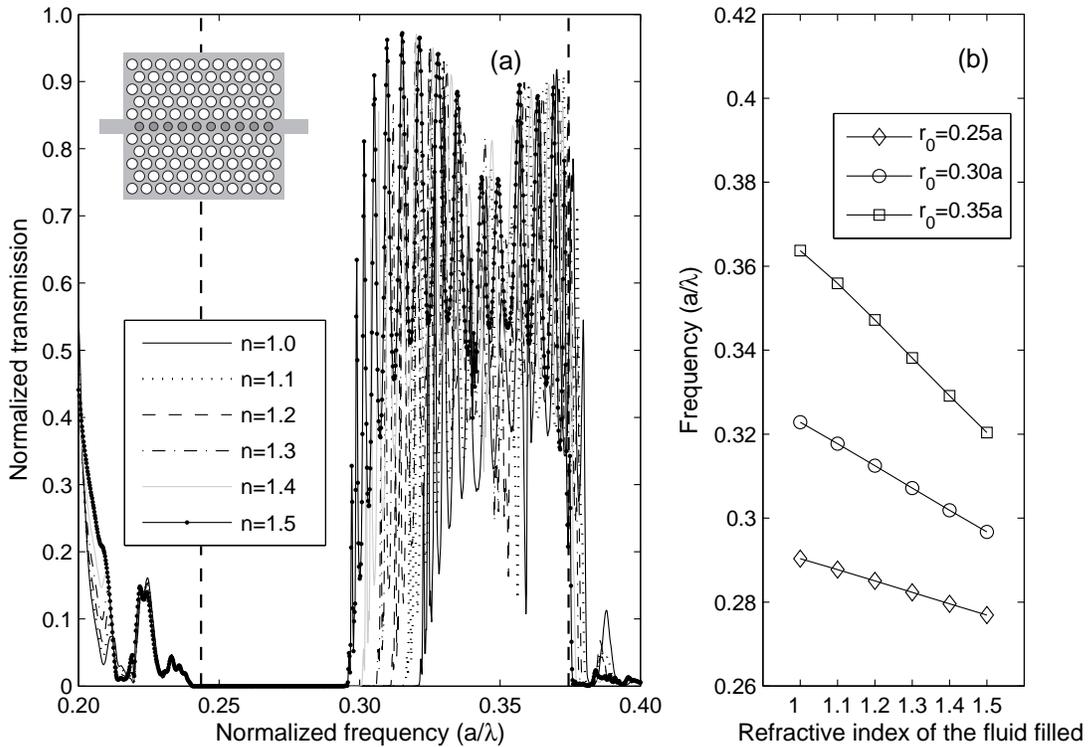}
\caption{(a)Transmission spectra for the PhC waveguide, see inset,
with the central air holes ($r_0$=0.30a) being filled by different
liquids with refractive indices varying from $n_l$=1.0 to 1.5 in
steps of $\delta n_l$=0.1. The two vertical dashed lines
illustrate the band-gap region for the complete PhC. (b) Mode-gap
edge as a function of the refractive index for the filled liquid.}
\label{Trans-R030}
\end{figure}

Let us first consider a photonic crystal waveguide shown in the
inset of Fig.~\ref{Trans-R030}(a). The PhC studied in this paper
is a two-dimensional PhC with a triangular array of air holes in
an InP/GaInAsP dielectric background. We assume that the
dielectric medium is non-absorbing, and has a constant index of
refraction n=3.24. The radius of the air holes is r=0.37a, where a
is the lattice constant. We calculate the photonic band structure
by a plane-wave method~\cite{Johnson:2001} and the results show
that the PhC possesses a wide photonic bandgap for TE polarization
(magnetic field parallel to the air holes) for the normalized
frequency a/$\lambda$ between 0.2450 and 0.3744. A simple way of
making a photonic crystal waveguide is to remove air holes within
a single row. However, in our case we need to realize a non-solid
waveguide serving also as a microfluidic channel or cavity. The
PhC waveguide is formed by slightly lowering the radius of the air
holes in a single row and these holes are used for local
refractive index modulation by selectively filling them with
liquid. Presently, the common techniques for local index
modulation are based on relatively weak nonlinearities, where
$\delta n/n$ is of the order of $10^{-3}$ or lower. Other
techniques are able to offer much higher $\delta n/n$, such as
mechanical deformation, thermo-optics, and liquid crystal
infusion. However, the effects tend to be of non-local nature.
Nanofluidics provides both localized control and high refractive
index modulation, which has attracted much attention. Recently,
Erickson {\it et al.} experimentally demonstrated nanofluidic
tuning of photonic crysal circuits~\cite{Erickson:2006}. Here, we
will focus our study on transmission spectra of the PhC waveguide,
shown in the inset of Fig.~\ref{Trans-R030}(a), for the waveguide
with air holes being filled with different liquids.
Transmission spectra for the PhC waveguide are obtained using the
finite-difference time-domain (FDTD) method~\cite{TafloveFDTD}
with the boundaries treated within the framework of perfectly
matched layers~\cite{Berenger:1994}. The in and out-coupling of
light is facilitated by index-guiding slab waveguides with a
TE-polarized fundamental mode as the source incident into the PhC
waveguide. The width of the input/output waveguide is $a$ and the
refractive index is equivalent to that of the background of the
PhC. It should be noted that the PhC waveguide may be multimode,
in which case only modes with a symmetry corresponding to the
source can be excited. Figure~\ref{Trans-R030}(a) shows
transmission spectra for the PhC waveguide with central air holes
being filled by different liquids with the refractive index
increasing from $n_l$=1.0 to $n_l$=1.5 in steps of $\delta
n_l$=0.1, where the radius of the central air holes is 0.30a. The
bandgap edge of the complete PhC is indicated by two vertical
dashed lines in Fig.~\ref{Trans-R030}(a). From
Fig.~\ref{Trans-R030}(a), one knows that light, with a frequency
in the bandgap, can propagate down the waveguide, and that there
exists a mode-gap region in the lower frequency region, where
propagation is inhibited. Although the PhC waveguide is a lossless
one, the transmission is never close to unity due to the coupling
loss at the two boundaries to the slab waveguides. Peaks in the
transmissions arise from the Fabry--Perot oscillations from the
two boundaries.

For the present application we are not interested in the details
of the Fabry--Perot pattern in Fig.~\ref{Trans-R030}(a), but
rather the spectral position of the mode-gap edge. As seen, the
low-frequency mode-gap edge does not change with the refractive
index of the liquid since it is a property of surrounding PhC. The
positions for the low-frequency mode-gap edge are in agreement
with that for band-gap edge indicated by the left vertical dashed
line in Fig.~\ref{Trans-R030}(a). However, the high-frequency
mode-gap edge is strongly dependent on the refractive index of the
liquid, as shown in Fig.~\ref{Trans-R030}(b) for the cases of
$r_0$=0.25a, 0.30a and 0.35a. As an example, for the case of
$r_0$=0.30a, the mode-gap edge changes from $a/\lambda$=0.322869
to 0.317751 when the air holes (with index $n_l$=1) are filled by
a liquid of index $n_l$=1.1. From Fig.~\ref{Trans-R030}(b), one
finds that the sensitivity becomes better as the hole size of the
central increases from $r_0$=0.25a to 0.35a. However, calculated
results show that the sensitivity does not depend much on the
length of the device when the length is larger than 7a. Now
consider a commercial silicone fluid with a calibrated
refractive-index accuracy of $\delta n_l$=0.002, as mentioned in
Ref.~\cite{Chow:2004}, where the refractive index of the liquid
varies from $n_l$=1.446 to 1.454 in increments of 0.002. For the
working wavelength around $1.55\rm {\mu} m$ (here we choose
a=450$\rm {\mu} m$), the mode-gap edge shifts up to 0.5~nm for
$\delta n_l$=0.002 when $r_0$=0.30a. For comparison, we note that
the shift in resonant wavelength for the high-quality-factor (Q)
PhC cavity is about 0.38 nm for $\delta n_l$=0.002 with a
discrepancy of 4\% between the calculated and experiment
results~\cite{Chow:2004}. The above results demonstrate that even
such a simple PhC waveguide has potential applications as a
sensitive sensor. Shifts in the frequency can be approximately
analyzed by first-order electromagnetic perturbation theory
yielding
\begin{eqnarray}
\frac{\delta \lambda}{\delta n_l} \approx \frac{\lambda}{n_l}f,
\label{eqn}
\end{eqnarray}
where $n_l$ is the refractive index of the liquid, $\lambda$ is
the working wavelength and $f$ is the filling factor of the energy
residing in the liquid, defined by
\begin{equation}
f=\frac{\int_l  d\vec{r}\, \vec{E}(\vec{r})\cdot
\vec{D}(\vec{r})}{\int d\vec{r}\, \vec{E}(\vec{r})\cdot
\vec{D}(\vec{r})}
\end{equation}
where $D=\epsilon E$ is the displacement field. The integral in
the numerator of the filling factor $f$ is restricted to the
region containing the fluid while the other integral is over all
space. For the low-frequency gap-edge mode mentioned above,
$\delta\omega$ is nearly zero since $f$ is close to zero due to
the bulk slab mode. $f$ is about $20\%$ for the high-frequency
gap-edge mode when $n_l$=1.446. Thus, the problem is really that
of maximizing the filling factor.

\begin{figure}[ht]
\centering\includegraphics[width=\columnwidth]{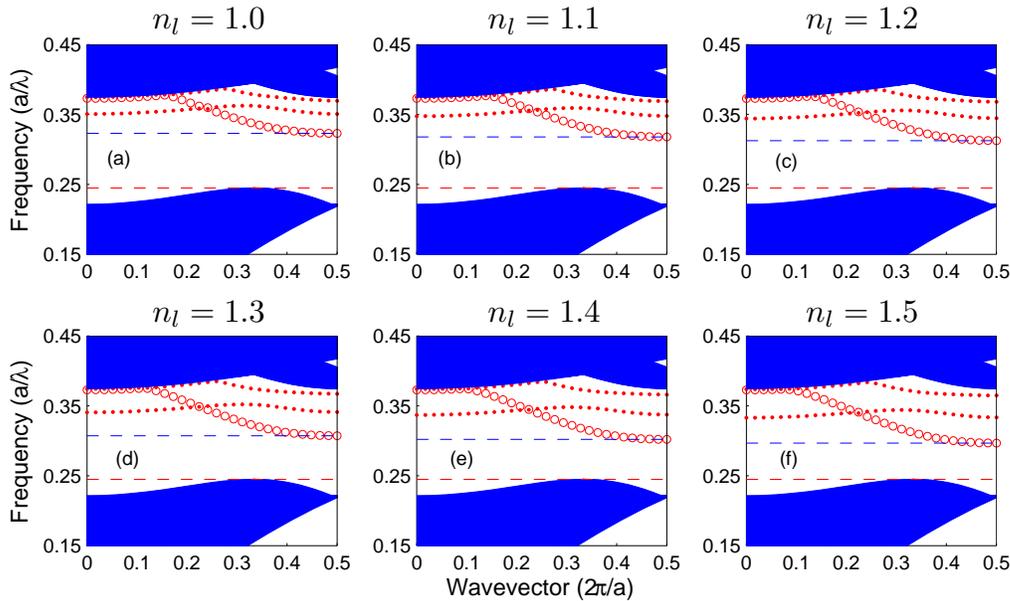}
\caption{(Color on line) Dispersion of the PhC air-holes waveguide
shown in the inset of Fig. 1(a), where the central air holes
($r_0$=0.30a) are filled by liquids with the refractive indices
varying in steps of $\delta n_l=0.1$. The circles and dots
represent the even and odd waveguide mode, respectively. }
\label{Band-R030}
\end{figure}

To further understand the physics behind it, we next support the
picture by dispersion calculation for the PhC waveguide. For this
purpose we use a plane-wave method~\cite{Johnson:2001}. The
dispersion of the PhC waveguide in absence or presence of a fluid
is shown in Fig.~\ref{Band-R030}, which clearly illustrates how a
waveguide mode forms within the bandgap region.
Figure~\ref{Band-R030} (a)-(f) summarize the dispersions for the
PhC waveguide, where the air holes, of radius $0.3a$, are filled
by a liquid with a varying refractive index. The shaded regions
are the projected band structure for TE slab modes, while the dots
and circles represent the odd and even waveguide modes,
respectively. The existence of a waveguide-mode gap is indicated
by two horizontal dashed lines, where the mode-gap edges are in
agreement with those obtained from the transmission spectra in
Fig.~\ref{Trans-R030}(a). When increasing the refractive index of
the liquid [going from panel (a) toward panel (f)], the
high-frequency mode-gap edge (blue-dashed line) is significantly
downward shifted. However, the low-frequency mode-gap edge,
related to the band-gap edge, does not change as the refractive
index of the liquid increases. We emphasize that all results
obtained from band structures are consistent with those from the
transmission spectra. The sensitivity of this structure is mainly
attributed to the dispersion of the even waveguide mode denoted by
circles in Fig.~\ref{Band-R030}.

\begin{figure}[ht]
\centering\includegraphics[width=\columnwidth]{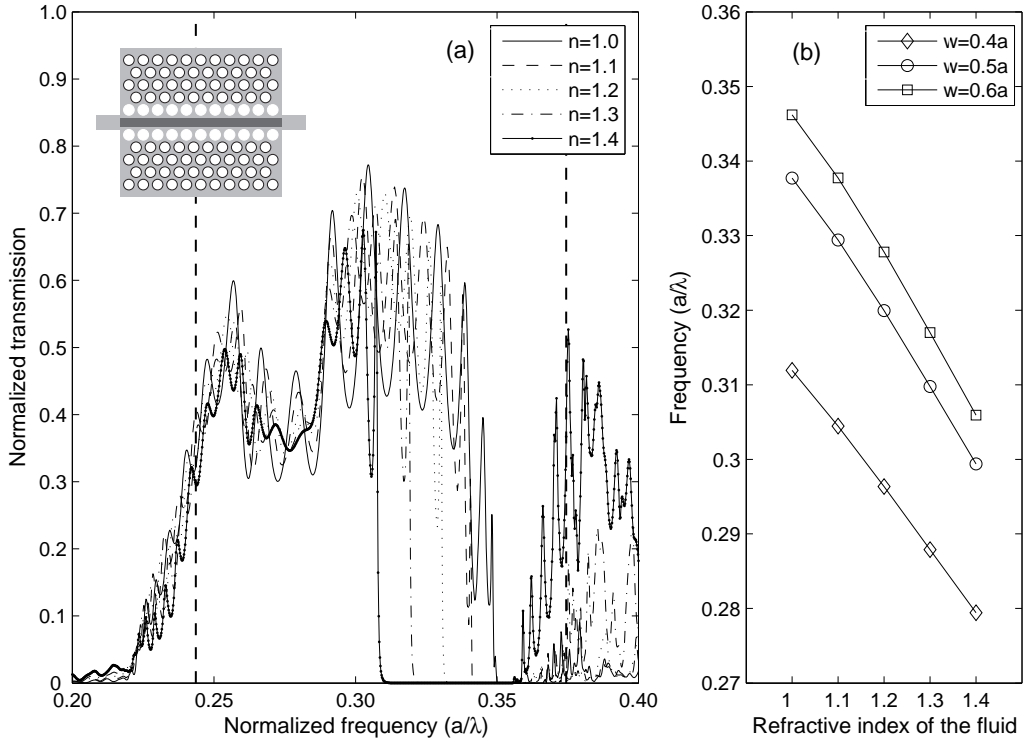}
\caption{(a)Transmission spectra for the PhC channel waveguide
(w=0.60a), see inset, being filled by different liquids. (b)
Low-frequency mode-gap edge as a function of the refractive index
for the filled liquid.} \label{Trans-D060}
\end{figure}

Let us next consider another photonic crystal waveguide shown in
the inset of Fig.~\ref{Trans-D060}(a), where the waveguide is
introduced by removing the background medium leaving us with an
air channel waveguide. The width of the air channel waveguide is
denoted by w, while other parameters and the input/output
waveguide are totally same as those mentioned above. Transmission
spectra, for the filled liquid with five different indices from
$n_l$=1.0 to $n_l$=1.4 in increments of $\delta n_l$=0.1, are
shown in Fig.~\ref{Trans-D060}(a) when w=0.6a. The band-gap region
is also indicated by two vertical dashed lines. Similar to the
result for the structure shown in Fig.~\ref{Trans-R030}(a), the
waveguide mode-gap region still exists. The high-frequency
mode-gap edge slightly changes as the refractive index of the
liquid increases, while the low-frequency mode-gap edge is
strongly dependant on the liquid. The low-frequency mode-gap edge
is downward shifted as the refractive of the liquid increases.
Here, we are only interested with the change of the position for
the slow-frequency mode-gap edge, as shown in
Fig.~\ref{Trans-D060}(b). One finds that the slope for the
sensitivity hardly change when varying the width of the channel
waveguide. Compared with the results for the high-frequency
mode-gap edge shown in Fig.~\ref{Trans-R030}(b), from
Fig.~\ref{Trans-D060}(b) one can find that such a waveguide shows
a better sensitivity. From Fig.~\ref{Trans-D060}(b) one also find
that sensitivity does not change much when varying the width of
the channel waveguide from w=0.4a to 0.6a. The mode-gap edge
shifts $\delta(a/\lambda)$=0.0085 when the air holes are filled by
a liquid of index $n_l$=1.1 for the case of w=0.6a. For comparison
we have $\delta(a/\lambda)=0.0051$ for the structure shown in the
inset of Fig.~\ref{Trans-R030}(a) when $r_0$=0.30a. The filling
factor of the low-frequency gap-edge mode is about 44\%, which is
higher than that of the high-frequency gap-edge mode for the first
structure shown in the inset of Fig.~\ref{Trans-R030}. Results
from the numerical calculations are consistent with those obtained
from the perturbation theory, Eq.(\ref{eqn}). Again we consider a
commercial silicone fluid with a calibrated refractive-index
accuracy of $\delta n_l$=0.002. For the working wavelength around
$1.55\,{\rm \mu m}$ (a=450$\,{\rm nm}$), the mode-gap edge shifts
up to 1.0~nm for $\delta n_l$=0.002.  The proposed sensor relies
strongly on the dispersion of the PhC waveguide mode and the
presence of a mode gap. To further improve the sensitivity, we
optimize the PhC waveguide structure by varying the radius ($r_1$)
of the first row of air holes surrounding the channel waveguide.
For simplicity we fix the width of the channel waveguide to 0.6 a.
We note that the device only works when the mode-gap exists and
that the mode-gap will disappear as $r_1$ increases. By a careful
design of the structure shown in the inset of
Fig.~\ref{Trans-D060}(a), we have been able to improve the design
further. For the working wavelength around $1.55\,{\rm \mu m}$
(a=450\,${\rm nm}$), the mode-gap edge shifts about 1.2~nm for
$\delta n_l=0.002$, when $r_1$ is tuned to 0.45 a. Besides, this
device has only a size about $5\,{\rm \mu m} \times 5\,{\rm \mu
m}$, which is sufficiently compact for most applications. However,
from Eq.(\ref{eqn}), one knows that the sensitivity is
proportional to the filling factor f for a specific working
wavelength. Since the filling factor may be close to unity,
photonic crystal devices may potentially have a sensitivity close
to the limit.

\section{Summary}
In this paper, we have theoretically studied optofluidic sensors
based on highly dispersive two-dimensional photonic crystal
waveguides. Our study shows that these structures are strongly
sensitive to the refractive index of the liquid, which is used to
tune dispersion of photonic crystal waveguides. For the working
wavelength around $1.55{\rm \mu m}$, waveguide mode-gap edge
shifts up to 1.2~nm in increments of $\delta n_l=0.002$. Compared
with the air-holes waveguide, the channel waveguide in the PhC
shows a better sensitivity, which can be explained by the
first-order perturbation theory. Our study shown above is based on
two-dimensional photonic crystal waveguides, while, it can be
easily extended to a two-dimensional photonic crystal slab. Even
though it may be difficult to inject liquid into the waveguides,
we would like to emphasize that recently Erickson {\it et al.}
experimentally demonstrated nanofluidic tuning in a similar
structure~\cite{Erickson:2006}. These devices show a potential for
chemical sensing applications.

\section*{Acknowledgments}
This work is financially supported by the \emph{Danish Council for
Strategic Research} through the \emph{Strategic Program for Young
Researchers} (grant no: 2117-05-0037).

\section*{References}
%\bibliographystyle{unsrt}
%\bibliography{OFTS}

%\section*{References}

\end{document}